\documentclass[prd,aps,nofootinbib,superscriptaddress,floatfix]{revtex4}
\pdfoutput=1

\usepackage{epsfig}
\usepackage{amsmath}
\usepackage{dsfont}

\input{epsf}

\begin{document}

\title{Quark orbital dynamics in the proton from Lattice QCD --
from Ji to Jaffe-Manohar orbital angular momentum}

\author{M.~Engelhardt \\
\it Department of Physics, New Mexico State University,
Las Cruces, NM 88003, USA}

\begin{abstract}
Given a Wigner distribution simultaneously characterizing quark
transverse positions and momenta in a proton, one can
directly evaluate their cross-product, i.e., quark orbital angular
momentum. The aforementioned distribution can be obtained by generalizing
the proton matrix elements of quark bilocal operators which define
transverse momentum-dependent parton distributions (TMDs); the transverse
momentum information is supplemented with transverse position information
by introducing an additional nonzero momentum transfer. A gauge connection
between the quarks must be specified in the quark bilocal operators;
the staple-shaped gauge link path used in TMD calculations yields the
Jaffe-Manohar definition of orbital angular momentum, whereas a straight
path yields the Ji definition. An exploratory lattice calculation, performed
at the pion mass $m_{\pi } = 518\, \mbox{MeV} $, is presented which
quasi-continuously interpolates between the two definitions and
demonstrates that their difference can be clearly resolved. The
resulting Ji orbital angular momentum is confronted with traditional
evaluations based on Ji's sum rule. Jaffe-Manohar orbital angular momentum
is enhanced in magnitude compared to its Ji counterpart.
\end{abstract}

\maketitle

% TODO ??????
% 
% discuss rogers

\section{Introduction}
Decomposing the proton spin into the spin and orbital angular momentum
contributions of its quark and gluon constituents has been a prominent
endeavor of hadronic physics since the groundbreaking EMC experiments
\cite{emc1,emc2} revealed that quark spin, by itself, does not account
for the entirety of proton spin. In particular, quantifying the parton
orbital dynamics already meets challenges at the conceptual level.
In a gauge theory, one cannot consider the quark degrees of freedom
in isolation; instead, quarks are intrinsically linked to gluonic
fields to conform to the strictures of gauge invariance. As a consequence,
there is no unique answer to the question of partitioning orbital angular
momentum \cite{leader,liulorce} into a quark and a gluonic contribution.
Quark orbital angular momentum will include gluonic effects to a varying
degree, depending on the chosen decomposition scheme.

Two definitions of quark orbital angular momentum have stood out in
particular. One is the Ji decomposition \cite{jidecomp}, which is
singled out by quark orbital angular momentum being defined in terms
of a quasi-local gauge-invariant operator. The other is the Jaffe-Manohar
decomposition \cite{jmdecomp}, which has the attribute of admitting
a partonic interpretation in the infinite momentum frame in the
light-cone gauge. Only fairly recently has a (non-local) gauge-invariant
extension been formulated for the Jaffe-Manohar definition of quark
orbital angular momentum \cite{hatta}, which opens the perspective
of calculating not only Ji, but also Jaffe-Manohar quark orbital
angular momentum in Lattice QCD. As will be discussed in detail below,
the difference between the two definitions can be encoded in differently
shaped gauge links in a nonlocal quark bilinear operator. In fact, by
deforming the gauge link in small steps between the two limits, it will
be possible to construct a gauge-invariant, quasi-continuous interpolation
between Ji and Jaffe-Manohar quark orbital angular momentum.

Hitherto, no direct evaluation of quark orbital angular momentum has
been available in Lattice QCD. Specifically Ji quark orbital angular
momentum has been calculated indirectly
\cite{liu1,liu2,reg1,reg2,reg3,etm1,etm2,LHPC_1,LHPC_2,LHPC_3}
as the difference between total angular momentum and spin, $L=J-S$,
taking advantage of Ji's sum rule \cite{jidecomp} relating $J$ to
generalized parton distributions (GPDs). Here, a first exploration of
a direct approach to calculating quark orbital angular momentum in
Lattice QCD is presented, based on a construction of a Wigner
distribution of quark positions and momenta in the proton. Utilizing
this Wigner distribution, average partonic orbital angular momentum
can be evaluated directly from the cross-product of position and momentum.
Ideas complementing the approach followed here have also been laid out
in \cite{yzhao,liuti}. The aforementioned Wigner distribution is derived
from the nonlocal quark bilinear operator mentioned further above, and
thus allows one to access a continuum of quark orbital angular momentum
definitions, including the Ji and Jaffe-Manohar ones, by varying the
gauge link between the quark operators.

Such Wigner distributions are related, through Fourier transformation,
to generalized transverse momentum-dependent parton distributions
(GTMDs) \cite{mms}, which differ from standard TMDs only by the
relevant proton matrix elements being evaluated at non-zero momentum
transfer. Thus, many of the concepts and techniques employed in the
following are derived from those used in lattice TMD studies
\cite{straightlett,straightlinks,tmdlat,bmlat}.

\section{Quark orbital angular momentum}
\label{oamsec}
In a longitudinally polarized proton propagating with large momentum
in the $3$-direction, the $3$-component of the orbital angular momentum
of quark partons can be evaluated directly in terms of their impact
parameter $b_T $ and transverse momentum $k_T $ given an appropriate
Wigner distribution of those quark characteristics \cite{lorce},
\begin{equation}
L^U_3 = \int dx \int d^2 b_T \int d^2 k_T \,
\left( b_T \times k_T \right)_3
{\cal W}^{U} ( x, b_T , k_T )
\label{oamwdef}
\end{equation}
where $x$ denotes the longitudinal quark momentum fraction. Note that,
since mutually orthogonal components of quark position and momentum
are being evaluated, no conflict with the uncertainty principle arises.
An appropriate Wigner distribution ${\cal W}^{U} ( x, b_T , k_T )$ is
given by the following Fourier transform with respect to the conjugate
pair of transverse vectors $b_T $ and $\Delta_{T} $,
\begin{equation}
{\cal W}^{U} ( x, b_T , k_T ) =
\int \frac{d^2 \Delta_{T} }{(2\pi )^{2} } e^{-i\Delta_{T} \cdot b_T }
\frac{1}{2} \left( W_{++}^U ( x, \Delta_{T} , k_T ) -
W_{--}^U ( x, \Delta_{T} , k_T ) \right)
\label{wignerft}
\end{equation}
in terms of the helicity-non-flip components of the correlator
$W_{\Lambda^{\prime } \Lambda }^U ( x, \Delta_{T} , k_T )$, which is of
the type that defines generalized transverse momentum-dependent parton
distributions (GTMDs) \cite{mms},
\begin{equation}
W_{\Lambda^{\prime } \Lambda }^U ( x, \Delta_{T} , k_T ) = \frac{1}{2}
\int \frac{d z^- d^2 z_T }{(2 \pi)^3} e^{i(xP^+ z^- - k_T \cdot z_T )}
\left. \frac{\langle p^{\prime } , \Lambda^{\prime } |
\overline{\psi}(-z/2) \gamma^+ U \psi(z/2)
| p, \Lambda \rangle }{ {\cal S} [U]} \right|_{z^+ =0} \ .
\label{gtmdcorr}
\end{equation}
Several comments are in order regarding this expression. In (\ref{wignerft})
and (\ref{gtmdcorr}), $\Delta_{T} $ denotes the
(transverse\footnote{Throughout this paper, the momentum transfer
to the proton will be taken to be purely transverse, i.e., the skewness
vanishes.}) momentum transfer to the proton, which carries momenta
$p=P-\Delta_{T} /2$, $p^{\prime } =P+\Delta_{T} /2 $ and helicities
$\Lambda $, $\Lambda^{\prime } $ in the initial and final states,
respectively; $P$ is in the 3-direction. This is a generalization
of the correlator used to define transverse momentum-dependent parton
distributions (TMDs) \cite{boer,collbook} to non-zero momentum
transfer; it thus includes information on the parton impact parameter
$b_T $, resolved via the Fourier transformation (\ref{wignerft}), in
addition to the parton momentum information encoded via the quark
operator separation $z$.

The correlator (\ref{gtmdcorr}) furthermore depends on the form of the
gauge link $U$ connecting the quark operator positions $-z/2 $ and $z/2$,
along with a soft factor ${\cal S} [U]$ required to regulate the divergences
associated with $U$. This soft factor is identical to the one used for
TMDs, since (\ref{gtmdcorr}) only differs from the standard TMD correlator
by the choice of external states, while retaining the same bilocal operator.
The divergences regulated by ${\cal S} [U]$ are associated with the latter,
not the former. This has been explored more concretely in \cite{gtmdsoft}.
In the development further below, appropriate ratios of correlators will be
considered in which the soft factors cancel, in analogy to previous lattice
TMD studies \cite{tmdlat,bmlat}; as a result, it is not necessary to specify
${\cal S} [U]$ in detail here, beyond a simple symmetry to be noted below.

\begin{figure}[h]
\centerline{\psfig{file=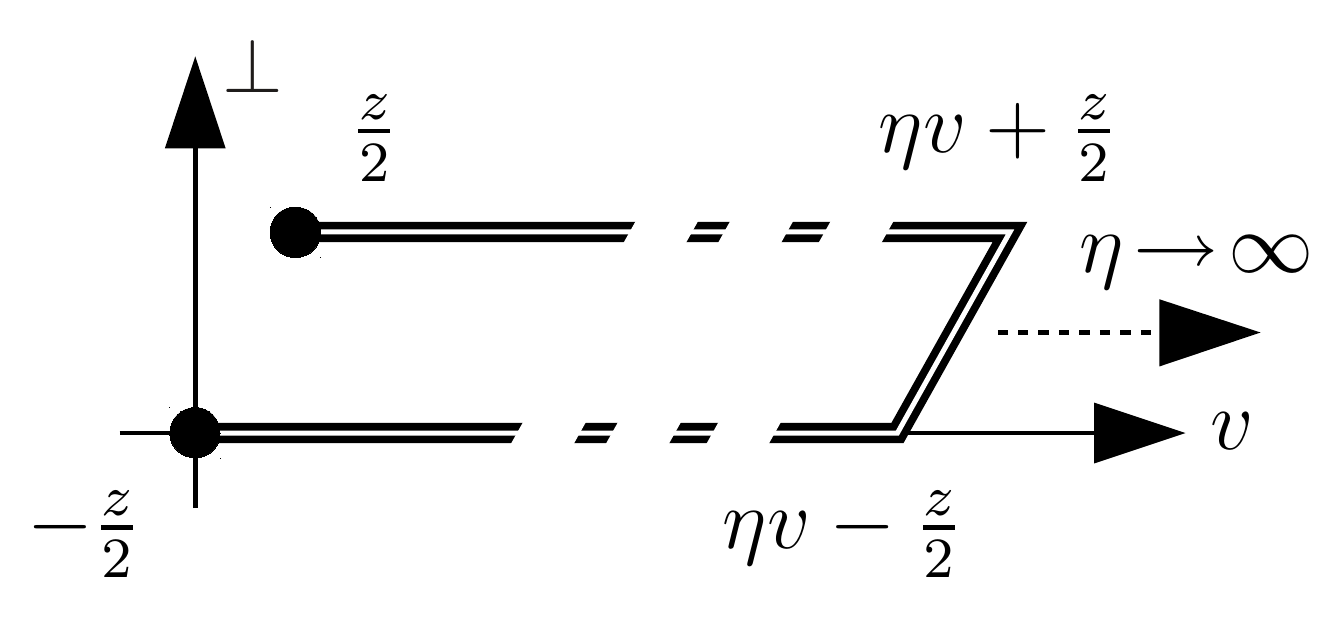,width=7cm} }
\caption{Path of the gauge connection $U$, cf.~(\ref{udef}), in the
correlator (\ref{gtmdcorr}).}
\label{staplefig}
\end{figure}

Two important concrete choices for the gauge link are a staple-shaped
gauge link path, as employed in the definition of TMDs, and a straight
gauge link path. A staple-shaped gauge link path is used, e.g., to encode
final state interactions of the struck quark in a deep-inelastic
scattering process, and yields Jaffe-Manohar quark orbital angular
momentum \cite{hatta,burk}; a straight gauge link path omits these
final state interactions, and yields Ji quark orbital angular momentum
\cite{jist,burk,eomlir}. The difference between the two definitions of
orbital angular momentum can thus be interpreted as the torque accumulated
owing to final state interactions experienced by a quark exiting the
proton after having been struck by a hard probe \cite{burk}. In the
lattice calculations pursued in the present work, staple-shaped gauge
link paths with varying staple lengths will be treated,
\begin{equation}
U\equiv U[-z/2,\eta v-z/2,\eta v+z/2,z/2]
\label{udef}
\end{equation}
where the gauge link connects the points given as the arguments of $U$
in (\ref{udef}) with straight Wilson lines, cf.~also Fig.~\ref{staplefig}.
The vector $v$ encodes the direction of the staple, the length of which
is scaled by the parameter $\eta $. For $\eta =0$, the gauge link
degenerates to a straight link connecting the points $-z/2$ and $z/2$
directly. In the notation adopted below, also negative values of
$\eta $ will be used, to denote a staple extending in the direction
opposite to $v$.
By allowing for continuous variation of $\eta $, the results obtained in
the present work will interpolate continuously and gauge-invariantly
between the $\eta =0$ Ji limit and the large $|\eta |$ Jaffe-Manohar
limit; note that quark orbital angular momentum is an even function
of $\eta $. In the convention adopted below, the $\eta \geq 0$ branch
will correspond to a struck quark gradually gathering up torque as it
is leaving the proton, owing to final state interactions.

The direction $v$ of the staple is associated with
the path of the struck quark in, e.g., a deep-inelastic scattering process;
in such a process, the Wilson lines forming the legs of the staple-shaped
gauge link represent eikonal approximations of gluon exchanges in the
final state. Here, the standard TMD specification \cite{ji04,aybat,collbook}
$v_T =0$ will be followed. Since such a hard process is characterized by a
large rapidity difference between the struck quark and the initial hadron,
a straightforward choice of $v$ which suggests itself would be a light-cone
vector. However, this choice leads to severe rapidity divergences
\cite{rapidrev}, which in the factorization framework developed in
\cite{aybat,collbook} are regulated by taking the staple direction $v$
off the light cone into the spacelike region. The correlator
(\ref{gtmdcorr}) can consequently be characterized in Lorentz-invariant
fashion by the additional Collins-Soper type parameter
\begin{equation}
\hat{\zeta } = \frac{v\cdot P}{\sqrt{|v^2 |} \sqrt{P^2 } }
\label{zetadef}
\end{equation}
in terms of which the light-cone limit is recovered for
$\hat{\zeta } \rightarrow \infty $. Note that the specification $v_T =0$
implies that the soft factor ${\cal S} [U]$ does not depend on the
direction of $z_T $, the transverse part of the quark operator separation
$z$; one can freely rotate $z_T $ in the transverse plane while keeping all
other geometric characteristics of the gauge link fixed. In the following,
the notation ${\cal S} [U] \equiv {\cal S} (z_T^2 )$ will be adopted to
emphasize this form of the dependence on $z_T $, suppressing dependences
on other aspects of the geometry of the gauge link $U$.

An important point to note is that adopting a scheme in which $v$ is chosen
to be spacelike is crucial to enable a Lattice QCD evaluation of the proton
matrix element in (\ref{gtmdcorr}). Since Lattice QCD employs a Euclidean
time dimension to project out the hadronic ground state, no physical temporal
extent can be accomodated in the quark bilocal operator of which one is
taking the matrix element. All separations must be purely spatial.
Only if $v$ is generically spacelike can one boost the problem to a
Lorentz frame in which $v$ becomes purely spatial, and perform the
calculation in that frame. Note that, in such a frame, large $\hat{\zeta } $
are realized by large spatial proton momenta ${\bf P} $, cf.~(\ref{zetadef}).
Specifically, in the calculations to follow, $v$ will point purely in
longitudinal direction, $v\equiv -\vec{e}_{3} $, and, consequently,
$\hat{\zeta } = P_3 /m$, with $m$ denoting the proton mass. Note
furthermore that, in the $\eta =0$ limit, the vector $v$ ceases to
play an independent physical role as the direction of propagation of
the struck quark, and thus $\hat{\zeta } $ ceases to encode a rapidity
difference between the proton and the struck quark. Nevertheless, in the
specific frame employed for the lattice calculation, one can still use
$v\equiv -\vec{e}_{3} $ to construct the parameter $\hat{\zeta } $, which
then merely continues to characterize the proton momentum in that frame.
Throughout the developments below, $\hat{\zeta } $ will consistently serve
to parametrize the approach to the physical limit. Of course, the
set of proton momenta which will be accessible in practice in the
concrete lattice calculation will be limited; as in lattice TMD
calculations \cite{tmdlat,bmlat}, extracting information about the large
$\hat{\zeta } $ regime presents a challenge in the present calculational
framework.

Inserting (\ref{wignerft}) into (\ref{oamwdef}), one obtains quark
orbital angular momentum in terms of the GTMD correlator (\ref{gtmdcorr}),
\begin{equation}
L^U_3 = -i\int dx \int d^2 k_T \,
\epsilon_{ij} k_{T,j}
\left. \frac{\partial }{\partial \Delta_{T,i} } \frac{1}{2}
\left( W_{++}^U ( x, \Delta_{T} , k_T ) -
W_{--}^U ( x, \Delta_{T} , k_T ) \right) \right|_{\Delta_{T} =0}
\label{lwppmm}
\end{equation}
Note that the GTMD correlator (\ref{gtmdcorr}) is parametrized by the
GTMDs $F_{1i} $, cf.~\cite{mms},
\begin{equation}
W_{\Lambda^{\prime } \Lambda }^U = \frac{1}{2m}
\bar{u} (p^{\prime } , \Lambda^{\prime } ) \left[
F_{11} + \frac{i\sigma^{i+} k_T^i }{P^{+} } F_{12} +
\frac{i\sigma^{i+} \Delta_{T}^{i} }{P^{+} } F_{13} +
\frac{i\sigma^{ij} k_T^i \Delta_{T}^{j} }{m^2 } F_{14} \right]
u(p,\Lambda )
\label{wparam}
\end{equation}
Inserted into the particular structure (\ref{lwppmm}), only the $F_{14} $
contribution survives, and one obtains an expression for quark orbital
momentum in terms of this GTMD \cite{lorce},
\begin{equation}
L^U_3 = -\int dx \int d^2 k_T \, \left. \frac{k_T^2 }{m^2 } F_{14}
\right|_{\Delta_{T} =0}
\label{lf14}
\end{equation}
Since it is cast purely in terms of momenta, this formulation is best
suited for phenomenological applications. The natural setting for a lattice
calculation, on the other hand, is not $k$-space, but instead $z$-space,
i.e., one calculates directly the proton matrix element in
(\ref{gtmdcorr}). Inserting (\ref{gtmdcorr}) into (\ref{lwppmm}),
quark orbital angular momentum takes the form
\begin{equation}
L^U_3 = \frac{1}{2P^+ } \epsilon_{ij} \frac{\partial }{\partial z_{T,i} }
\frac{\partial }{\partial \Delta_{T,j} } \!
\left. \frac{1}{2}
\frac{\langle p', + | \overline{\psi} (-z/2) \gamma^+ U \psi(z/2)
| p, + \rangle -\langle p', - | \overline{\psi} (-z/2) \gamma^+ U \psi(z/2)
| p, - \rangle }{ {\cal S} (z_T^2 )}
\right|_{z^+ = z^- =0\, , \ \Delta_{T} =0\, , \ z_T \rightarrow 0}
\label{ldercomb}
\end{equation}
Thus, in terms of the $z$ and $\Delta $ variables preferred for a Lattice QCD
calculation, no Fourier integrals over the generated data are necessary to
extract quark orbital angular momentum; this significantly facilitates
the numerical analysis. The expression (\ref{ldercomb}) can still be
brought into a more compact form. Owing to the symmetric treatment of
both the momentum transfer $\Delta_{T} $ and the spin, replacing the
proton helicities in (\ref{ldercomb}) by proton spins pointing into the
fixed positive and negative $3$-directions, respectively, only generates
correction terms in (\ref{ldercomb}) which vanish as
$\Delta_{T} \rightarrow 0$. Secondly, the two terms in
(\ref{ldercomb}) then yield identical contributions to quark
orbital angular momentum, and one can therefore confine oneself to
evaluating either one of the two terms. Thus, quark orbital angular
momentum can be evaluated more simply as
\begin{equation}
L^U_3 = \frac{1}{2P^+ }
\epsilon_{ij} \frac{\partial }{\partial z_{T,i} }
\left. \frac{\partial }{\partial \Delta_{T,j} }
\frac{\langle p^{\prime } , S^{\prime } =\vec{e}_{3} |
\overline{\psi} (-z/2) \gamma^+ U \psi(z/2)
| p, S=\vec{e}_{3} \rangle }{ {\cal S} (z_T^2 )}
\right|_{z^+ = z^- =0\, , \ \Delta_{T} =0\, , \ z_T \rightarrow 0}
\label{ldersingle}
\end{equation}
Note that, formally, one might consider taking the soft factor outside
the $z_T $-derivative in the limit $z_T \rightarrow 0$, since the soft
factor depends only on the magnitude of $z_T $, not its direction. However,
in the presence of a lattice cutoff, the derivative with respect to $z_T $
will be realized as a finite difference, and the limit $z_T \rightarrow 0$
has to be taken with care. For this reason, the soft factor is left inside
the derivative at this stage, deferring a more precise treatment to the
practical implementation of (\ref{ldersingle}) in the presence of a
lattice cutoff.

\section{Regularization and renormalization}
The form (\ref{ldersingle}) for quark orbital angular momentum is not
directly amenable to lattice evaluation, because the soft factor in the
Collins factorization scheme \cite{aybat,collbook,spl2} is not accessible
in Lattice QCD; it contains Wilson line structures of varying rapidities,
not all of which can be simultaneously boosted to a single instant, a
prerequisite for lattice evaluation\footnote{An exception is the
$\eta =0$ limit, in which the staple direction vector $v$ ceases to
play a physical role; this case was discussed in detail in
\cite{straightlinks}.}. As in previous lattice TMD studies
\cite{tmdlat,bmlat}, this is addressed by forming an appropriate ratio
with an additional quantity, such that the soft factors cancel. A convenient
choice of such a second quantity can be constructed via the unpolarized
TMD $f_1 (x,k_T )$. This TMD is obtained from the GTMD $F_{11} $ in the
limit of vanishing momentum transfer \cite{mms}, and thus from the
correlator (\ref{gtmdcorr}) as, cf.~the parametrization (\ref{wparam}),
\begin{equation}
f_1 = \left. F_{11} \right|_{\Delta_{T} =0}
= \left. W^{U}_{++} \right|_{\Delta_{T} =0}
\end{equation}
($W^{U}_{--} $ could likewise be used). On the other hand, integrating
$f_1 $ over all momenta yields the number of valence quarks,
\begin{equation}
n = \int dx \int d^2 k_T \, f_1 =
\frac{1}{2P^{+} }
\left. \frac{\langle p^{\prime } , S^{\prime } =\vec{e}_{3} |
\overline{\psi}(-z/2) \gamma^+ U \psi(z/2)
| p, S =\vec{e}_{3} \rangle }{ {\cal S} (z_T^2 )}
\right|_{z^+ =z^- =0,\Delta_{T} =0,z_T \rightarrow 0}
\label{ndenom}
\end{equation}
having inserted the explicit form of $W^{U}_{++} $, cf.~(\ref{gtmdcorr}),
and also having used that, for $\Delta_{T} =0$, positive helicity
corresponds to spin in the $3$-direction. Now, comparing (\ref{ndenom})
with (\ref{ldersingle}), the two matrix elements contain the same
operator, associated accordingly with the same soft factor; one can
therefore envisage canceling the soft factors by taking an appropriate
ratio. Care has to be taken, however, in defining the $z_T \rightarrow 0$
limit. At a finite lattice spacing $a$, distances smaller than $a$
cannot be resolved, and the $z_T $-derivative in (\ref{ldersingle})
is realized as a finite difference. It therefore necessarily involves
values of the correlator (\ref{ldersingle}) at finite $z_T $ for any
given finite lattice spacing $a$. Denoting
\begin{equation}
\Phi (z_T) = \langle p^{\prime } , S^{\prime } =\vec{e}_{3} |
\overline{\psi}(-z/2) \gamma^+ U \psi(z/2) | p, S =\vec{e}_{3} \rangle \ ,
\label{phicorr}
\end{equation}
(where, for conciseness of notation, all other dependences are being
omitted in the argument of $\Phi $), one can specify
\begin{equation}
L^U_3 = \frac{1}{2P^{+} } \epsilon_{ij}
\frac{\partial }{\partial \Delta_{T,j} } \frac{1}{2da}
\left. \frac{1}{ {\cal S} ((da)^2 ) }
\left( \Phi (da\vec{e}_{i} ) - \Phi (-da\vec{e}_{i} ) \right)
\right|_{z^+ = z^- =0,\Delta_{T} =0}
\label{ldiscrete}
\end{equation}
with a fixed number of lattice spacings $d$, which approximates the
derivative as $a\rightarrow 0$. One may contemplate using $d>1$ in
order to mitigate lattice artefacts; this will be revisited in the
discussion of numerical results below. Note that, since the soft
factor does not depend on the direction of $z_T $, but only its magnitude,
the two terms in the finite difference are associated with an identical
soft factor, which can therefore be factored out.

In order to use (\ref{ndenom}) to form a ratio in which the soft factor
cancels, one has to accordingly specify the limit $z_T \rightarrow 0$ as
\begin{equation}
n = \frac{1}{2P^{+} }
\left. \frac{1}{ {\cal S} ((da)^2 ) }
\frac{1}{2} \left( \Phi (da\vec{e}_{i} ) + \Phi (-da\vec{e}_{i} ) \right)
\right|_{z^+ = z^- =0,\Delta_{T} =0}
\label{ndiscrete}
\end{equation}
which is independent of the transverse direction $\vec{e}_{i} $ used and
approximates (\ref{ndenom}) as $a\rightarrow 0$ in a way which matches the
manner in which (\ref{ldiscrete}) approximates (\ref{ldersingle}), i.e.,
using values of $\Phi (z_T )$ at finite $|z_T | = da$ for any given finite
lattice spacing $a$. Combining (\ref{ldiscrete}) and (\ref{ndiscrete}),
one obtains the renormalized ratio
\begin{equation}
\frac{L^U_3 }{n} = \frac{1}{da} \epsilon_{ij}
\left. \frac{\frac{\partial }{\partial \Delta_{T,j} }
\left( \Phi (da\vec{e}_{i} ) - \Phi (-da\vec{e}_{i} ) \right) }{
\Phi (da\vec{e}_{i} ) + \Phi (-da\vec{e}_{i} )}
\right|_{z^+ = z^- =0,\Delta_{T} =0}
\label{rdiscrete}
\end{equation}
(where sums over $i,j$ are still implied), in which the soft factors,
and also any multiplicative renormalization constants attached to the
quark operators in $\Phi (z_T )$ have canceled. The ratio
(\ref{rdiscrete}) is the expression evaluated in practice in
the Lattice QCD calculation to be discussed below. Note that also
the derivative with respect to $\Delta_{T} $ in practice will have
to be approximated by a finite difference. However, contrary to
the treatment of the $z_T \rightarrow 0$ limit, the
$\Delta_{T} \rightarrow 0$ limit is not associated with ultraviolet
divergences and is conceptually straightforward, even if it poses
numerical challenges, to be discussed further below.

Besides considering the ratio (\ref{rdiscrete}) for fixed $d$, such that
$|z_T | = da \rightarrow 0$ for $a\rightarrow 0$, one can also
consider the regime of fixed finite physical distance $|z_T |=da$
and define a generalized, $|z_T |$-dependent quark orbital angular momentum
\begin{equation}
\frac{L^U_3 }{n} (|z_T |) = \frac{1}{|z_T |} \epsilon_{ij}
\left. \frac{\frac{\partial }{\partial \Delta_{T,j} }
\left( \Phi (|z_T |\vec{e}_{i} ) - \Phi (-|z_T |\vec{e}_{i} ) \right) }{
\Phi (|z_T |\vec{e}_{i} ) + \Phi (-|z_T |\vec{e}_{i} )}
\right|_{z^+ = z^- =0,\Delta_{T} =0}
\label{rdisczt}
\end{equation}
in analogy to, e.g., the generalized tensor charge defined in
\cite{tmdlat}. Keeping $|z_T |$ finite effectively acts as a
regulator on $k_T $-integrations such as the ones in (\ref{oamwdef})
and (\ref{lf14}).

As the quark operators in $\Phi (z_T )$, eq.~(\ref{phicorr}), approach
each other, i.e., in the limit $z_T \rightarrow 0$ once one has set
$z^+ = z^- =0$, the correlator becomes singular. In particular, even if
one has properly renormalized $\Phi $, at face value, its derivative
$\partial \Phi / \partial z_{T,i} $ in general will formally still be
divergent at $z_T =0$; in $k_T $-space, this translates to the
observation that, even if one has properly defined the GTMD $F_{14} $,
cf.~(\ref{wparam}),(\ref{lf14}), its $k_T $-moment (\ref{lf14}) still
exhibits an ultraviolet divergence. However, at any given resolution scale,
differences in $z_T $ smaller than that scale are not resolvable, and thus
a consistent interpretation of the $z_T $-derivative is in terms of
a finite difference taken over a distance commensurate with the
resolution scale; the aforementioned singularity occurring at arbitrarily
small distances is not resolved. As one increases the resolution, probing
the ultraviolet behavior more fully, one not only evaluates the
$z_T $-derivative using smaller distances; one concomitantly evolves
$\Phi $ to the higher resolution scale. The combined effect of these
adjustments ultimately produces the proper evolution of the orbital
angular momentum $L^U_3 $. An explicit observation of this evolution will
not be possible in the present work, since calculations were performed
only at one fixed lattice spacing $a$; it would be interesting to pursue
an evolution study in future work employing data at a sequence of
lattice spacings $a$.

In this context, also further remarks regarding the multiplicative nature
of the soft factors and the quark field renormalizations, and their
consequent cancellation in the ratio (\ref{rdiscrete}), are in order.
The absorption of divergences into these multiplicative factors is,
to begin with, a construction motivated by the continuum theory
\cite{aybat,collbook,spl2}. That it carries over analogously into the
lattice formulation is a working assumption which has been discussed in
some detail in \cite{straightlinks}, and was further explored empirically
in \cite{latt15_by} by investigating whether TMD ratios vary under changes
of the lattice discretization scheme. Absence of such variations
strengthens the hypothesis that the data provide a good representation
of the universal continuum behavior within statistical accuracy;
conversely, violations of the multiplicative nature of renormalization
would presumably manifest themselves in deviations from a universal
continuum limit. On the one hand, at finite physical separations $z_T $,
one would expect that the lattice operators approximate the continuum
operators well and inherit their divergence structure; indeed,
in \cite{latt15_by}, results for TMD ratios obtained using
differing discretization schemes were consistently found to coincide
once $z_T $ extends over several lattice spacings.
On the other hand, the limit of small $z_T $ deserves further
scrutiny. As already noted above, the $z_T \rightarrow 0$ limit,
in particular the $z_T $-derivative $\partial \Phi / \partial z_{T,i} $
in that limit, in general contains additional divergences; using
different lattice discretization schemes will have an effect analogous
to using different combinations of data obtained at varying $d$ to
construct the $z_T $-derivative, cf.~(\ref{ldiscrete}). Moreover, mixing
between the various composite local operators arising in the
$z_T \rightarrow 0$ limit can occur, potentially modifying the
simple multiplicative renormalization pattern. In the empirical study
\cite{latt15_by}, the stability of TMD ratios under changes of the lattice
discretization scheme was generally seen to persist into the small $z_T $
regime, indicating that the data continue to represent the universal
continuum limit well, with the notable exception of the worm-gear shift
$\langle {\bf k}_{x} \rangle^{g_{1T} } $, for which significant deviations
between different discretization schemes were observed. The Sivers shift
$\langle {\bf k}_{y} \rangle^{\mbox{\scriptsize Sivers} } $, which is the
TMD ratio most directly related to the ratio (\ref{rdiscrete}) considered
here, exhibited no such deviations. Apart from performing
an analogous study also directly for the ratio
(\ref{rdiscrete}), another opportunity to cross-check the behavior of
(\ref{rdiscrete}) for the need for additional renormalization will
ultimately be provided by a comparison of the $\eta =0$ Ji limit with
the result for quark orbital angular momentum obtained using Ji's sum rule,
on the same lattice ensemble. It should be noted, however, that the
numerical data generated to date will not yet permit definite conclusions
in this regard; at the present stage, the aforementioned comparison is
affected by significant other systematic uncertainties as well, as
discussed further below.

\section{Lattice calculation}
Numerical data for the present study were generated employing a
mixed action scheme in which the valence quarks are realized as
domain wall fermions, propagating on a $N_f =2+1$ dynamical asqtad quark
gauge background. The corresponding gauge ensemble was provided by
the MILC Collaboration \cite{milc}. A fairly high pion mass,
$m_{\pi}=518\, \mbox{MeV} $, was used for this exploration in order
to reduce statistical fluctuations. Table~\ref{Tab_1} lists the details
of the ensemble.
\begin{table}
\centering
\begin{tabular}{|c|c|c|c|c|c|c|c|}
\hline
$L^3\times T$ & $a$(fm) & $am_{u,d}$ & $am_s$ &
$m_{\pi}^{\mbox{\tiny DWF} }$ (MeV) & $m_N^{\mbox{\tiny DWF} } $ (GeV) &
\#conf. & \#meas. \\
\hline
$20^3\times 64$ & 0.11849(14)(99) & 0.02  & 0.05 &
518.4(07)(49) & 1.348(09)(13) & 486 & 3888 \\
\hline
\end{tabular}
\caption{Details of the lattice ensemble. The lattice spacing $a$ was
determined using a different scheme, cf.~\cite{tmdlat}, than was employed
in \cite{LHPC_1,LHPC_2}; as a result, it and the listed hadron masses
deviate slightly from the values quoted in the latter references.
The uncertainties given for the hadron masses are, in that order,
statistical, and stemming from the conversion to physical units via
$a$. The values $m_{u,d,s} $ are the bare asqtad quark masses. On
each gauge configuration, eight measurements were performed.}
\label{Tab_1}
\end{table}
The mixed action scheme employed here has been the basis for detailed
investigations into hadron structure carried out by the LHP Collaboration
\cite{LHPC_1,LHPC_2}, including the particular ensemble described in
Table~\ref{Tab_1}. It was also used in previous lattice TMD studies
\cite{tmdlat,bmlat}.

Evaluation of the correlator (\ref{phicorr}), needed to obtain quark
orbital angular momentum through (\ref{rdiscrete}), proceeds
via the calculation of appropriate three-point functions
$C_{\mbox{\tiny 3pt} } [\hat{O} ]$ as well as two-point functions
$C_{\mbox{\tiny 2pt} } $, projected onto a definite proton momentum
at the proton sink, as well as a definite momentum transfer at the
operator insertion in $C_{\mbox{\tiny 3pt} } [\hat{O} ]$,
\begin{eqnarray}
C_{\mbox{\tiny 3pt} } [\hat{O} ] (t,t_f ,p^{\prime }, p) &=&
\sum_{{\bf x}_{f} , \, {\bf y} }
e^{-i{\bf x}_{f} \cdot {\bf p^{\prime } }
+i {\bf y} \cdot ({\bf p^{\prime } -p} ) }
\mbox{tr} [\Gamma_{\mbox{\tiny pol} }
\langle n (t_f ,{\bf x}_{f} ) \hat{O} (t,{\bf y} )
\bar{n} (0,0) \rangle ] \\
C_{\mbox{\tiny 2pt} } (t_f ,p^{\prime } ) &=&
\sum_{{\bf x}_{f} }
e^{-i{\bf x}_{f} \cdot {\bf p^{\prime } } }
\mbox{tr} [\Gamma_{\mbox{\tiny pol} }
\langle n (t_f ,{\bf x}_{f} )
\bar{n} (0,0) \rangle ] \ ,
\end{eqnarray}
constructed using Wuppertal-smeared proton interpolating fields
$n(t,{\bf x} )$, with the projector $\Gamma_{\mbox{\tiny pol} } =
\frac{1}{2} (1+\gamma_{4} ) \frac{1}{2} (1-i\gamma_{3} \gamma_{5} ) $
selecting states polarized in the 3-direction. The Euclidean temporal
separation between proton sources and sinks was $t_f =9a$.
The operator $\hat{O} $ is taken to be the operator
in (\ref{phicorr}); as already discussed following eq.~(\ref{zetadef})
further above, the lattice calculation is carried out in a Lorentz frame
in which $\hat{O} $ exists at a single time $t$, i.e., the quark
operator separation $z$ and the staple direction $v$ are purely spatial.
In this frame, $v=-\vec{e}_{3} $ and, consequently, $\hat{\zeta } =P_3 /m$,
cf.~(\ref{zetadef}), in terms of which the physical limit is approached
as $\hat{\zeta } $ becomes large. In the present study, only connected
diagrams entering $C_{\mbox{\tiny 3pt} } [\hat{O} ]$ were calculated.
Computationally significantly more expensive disconnected contractions,
the magnitudes of which are expected to be minor at the heavy pion mass
employed here, are excluded from all numerical results quoted below
(they cancel in the isovector $u-d$ quark channel).

To obtain the correlator (\ref{phicorr}), one forms the three-point to
two-point function ratio
\begin{equation}
2 E(p^{\prime } )
\frac{C_{\mbox{\tiny 3pt} } [\hat{O} ]
(t,t_f ,p^{\prime } ,p)}{C_{\mbox{\tiny 2pt} } (t_f ,p^{\prime } )}
\longrightarrow \Phi (z_T )
\label{3to2rat}
\end{equation}
which exhibits plateaus in $t$ yielding $\Phi (z_T )$ for $0 \ll t \ll t_f $.
In (\ref{3to2rat}), $E(p^{\prime } )=E(p)$ is the energy of the final and
of the initial proton state; note that the symmetric treatment of the initial
and final momenta $p=P-\Delta_{T} /2$, $p^{\prime } =P+\Delta_{T} /2 $
simplifies the ratio required to extract $\Phi (z_T ) $ in (\ref{3to2rat})
compared to the case of general $p,p^{\prime } $ \cite{LHPC_1,LHPC_2}. In
the present exploration, possible excited state contaminations affecting the
plateau values determined from (\ref{3to2rat}) at the source-sink separation
$t_f = 9a = 1.07\, \mbox{fm} $ were not quantified. They are expected
to be small at the heavy pion mass $m_{\pi } =518\, \mbox{MeV} $ considered
here; however, it will be necessary to account for them in more quantitative
studies at lower pion masses.

In the mixed action calculational scheme employed in the present
investigation, HYP-smearing is applied to the gauge configurations
before computing the domain wall valence quark propagators. This
suppresses dislocations in the fields that could potentially
produce spurious mixing between the right-handed and left-handed
fermion modes localized on their respective domain walls. These
same HYP-smeared gauge fields were also used to assemble the gauge link
$U$ in (\ref{phicorr}). As a consequence, even before cancellation in
the ratio (\ref{rdiscrete}), renormalization constants and soft factors
correspond more closely to their tree-level values.

\begin{table}
\centering
\begin{tabular}{|c|c|c|c|}
\hline
${\bf P} \cdot aL/(2\pi )$ & ${\bf \Delta } \cdot aL/(4\pi )$
& ${\bf z}/a$ & $\eta {\bf v}/a$ \\
\hline
$(0,0,n_P )$ & $(n_{\Delta } ,0,0)$ & $(0,n_z ,0)$ & $(0,0,n_v )$ \\
\hline
$(0,0,n_P )$ & $(0,n_{\Delta } ,0)$ & $(n_z ,0,0)$ & $(0,0,n_v )$ \\
\hline
\end{tabular}
\caption{Combinations of momenta and gauge link geometries used. The
longitudinal momentum index $n_P $ took the values $n_P =0,1,2$; the
transverse momentum transfer index $n_{\Delta } $ took the values
$n_{\Delta } =-1,0,1$. Note that the symmetric treatment of the
momentum transfer in the initial and final states,
$p=P-\Delta_{T} /2$, $p^{\prime } =P+\Delta_{T} /2 $, implies
that, for the standard periodic spatial boundary conditions used in
the present study, the smallest nonzero momentum transfer available
is $4\pi /(aL)$. The quark separation index $n_z $, corresponding
to the width of the staple-shaped gauge link, extended over the integers
$n_z =-5\ldots 5$; the range of the index $n_v $ determining the staple
length of the gauge link covered all values for which a numerical signal
is discernible (in both the positive and negative directions).}
\label{Tab_2}
\end{table}

The combinations of gauge link geometries and momenta considered in the
calculation are given in Table~\ref{Tab_2}. As far as the gauge link
geometries are concerned, the focus of the present investigation is on
small values of the quark separation $z_T $, and therefore only the
restricted range of $n_z =-5\ldots 5$ in Table~\ref{Tab_2} was treated;
it is possible to obtain a numerical signal at larger $z_T $, cf.~the
TMD investigations \cite{tmdlat,bmlat}. On the other hand, the staple
length index $n_v $ covered all values for which a numerical signal is
discernible. In particular, for $n_v =0$, one accesses Ji quark
orbital angular momentum, whereas for asymptotically large $n_v $,
Jaffe-Manohar quark orbital angular momentum is probed. The correlators
are time-reversal even, i.e., invariant under reflection of $n_v $.

On the other hand, the set of proton momenta accessed in this exploratory
investigation is rather limited, both in terms of the longitudinal
momentum $P$ as well as the transverse momentum transfer $\Delta_{T} $.
This is the source of the most significant systematic uncertainties
affecting the results extracted, and thus represents the most severe
shortcoming of the data set collected for this study. To extract quark
orbital angular momentum, the longitudinal momentum ${\bf P}$ must be
extrapolated to large values. The spatial momenta employed in practice,
cf.~Table~\ref{Tab_2}, have magnitudes $|{\bf P }| = 0, 0.52, 1.04$ GeV
and thus remain far from the asymptotic regime. Simultaneously, in terms of
the Collins-Soper parameter $\hat{\zeta } $, characterizing the rapidity
difference between the vectors $P$ and $v$, these momenta correspond to
$\hat{\zeta } =0, 0.39, 0.78$, respectively; these values likewise
are still far from the region in which one would expect to be able
to connect to perturbative evolution in $\hat{\zeta } $. To give
a tentative indication of the results for quark orbital angular momentum
one might expect to obtain at large proton momenta, below, an ad hoc
extrapolation will be entertained using the available two nonvanishing
values of $|{\bf P }|$. This extrapolation will be guided by the
behavior that was observed in a study \cite{bmlat} of the
Boer-Mulders effect in a pion, in which it proved possible to extract
the large-$\hat{\zeta } $ limit. However, in order to obtain stringent
quantitative results for quark orbital momentum in the proton, it will
be necessary to generate data at larger $|{\bf P }|$, and, concomitantly,
larger $\hat{\zeta } $ in future work.

Additionally, to extract quark orbital angular momentum,
cf.~(\ref{rdiscrete}), one has to evaluate a derivative of the correlator
data with respect to the momentum transfer $\Delta_{T} $ in the forward
limit. To obtain an accurate estimate of this derivative, one needs to
employ data at small $\Delta_{T} $. However, the nonzero values of the
momentum transfer available for this study, cf.~Table~\ref{Tab_2}, are
quite substantial, $|\Delta_{T} | =1.04$ GeV; note that the symmetric
treatment of the momentum transfer in the initial and final states,
$p=P-\Delta_{T} /2$, $p^{\prime } =P+\Delta_{T} /2 $,
implies that, for the standard periodic spatial boundary
conditions used in the present study, the smallest nonzero momentum
transfer available is $4\pi /(aL)$. In practice, the numerical results
presented in the following were extracted by averaging over the finite
differences obtained in opposite directions, i.e., using directly the
difference between the $n_{\Delta } =\pm 1$ data; the $n_{\Delta } =0$
data served to evaluate the denominator in (\ref{rdiscrete}).
Using such a finite difference of correlator data at the aforementioned
large values of the momentum transfer will lead to a
substantial underestimate of the derivative. To obtain a rough idea
of the magnitude of the underestimate, note that the combination
$F=( \Phi (da\vec{e}_{i} ) - \Phi (-da\vec{e}_{i} ) )$ appearing in
the numerator of (\ref{rdiscrete}), of which one requires the
$\Delta_{T} $-derivative, is an odd function of $\Delta_{T} $.
Absent more detailed information on the form of its
$\Delta_{T} $-dependence, consider modeling it as
$F(\Delta_{T} ) = \Delta_{T} F_1 ( |\Delta_{T} |)$, using the isovector
electromagnetic Dirac form factor $F_1 $ as a proxy for the typical
variation of distribution functions with $\Delta_{T} $. The fall-off
of $F_1 $ with $\Delta_{T} $ then gives an indication of the extent
to which the numerator of (\ref{rdiscrete}) may be underestimated.
On the ensemble used in the present study, $F_1 $ falls off \cite{LHPC_2}
by (slightly more than) a factor 2 between $\Delta_{T} =0$ and
$|\Delta_{T} | = 1$ GeV. Thus, a substantial upward correction by a
factor of this magnitude must be expected for the extracted values
of the quark orbital angular momentum, once this source of systematic
bias is eliminated. This is the top priority for future more
quantitative studies, and a calculation is currently in preparation
which will incorporate a direct method to evaluate the
$\Delta_{T} $-derivative exactly \cite{rome,nhasan,smpriv}.

\section{Numerical results}
\begin{figure}
\centerline{\psfig{file=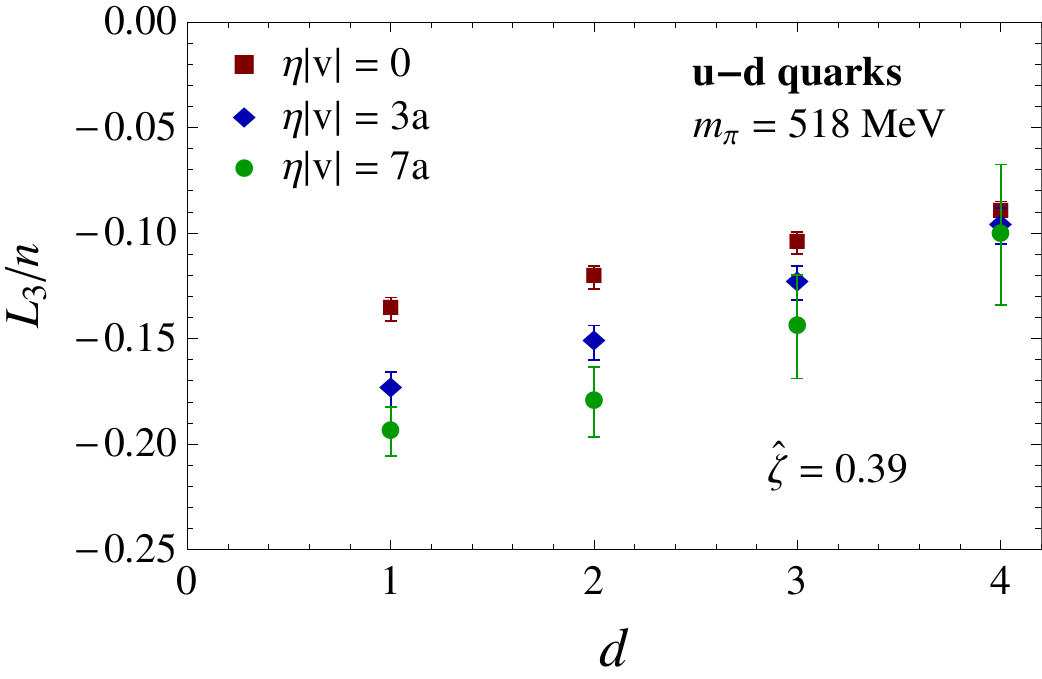,width=8.5cm} }
\caption{
Quark orbital angular momentum in units of the number of valence quarks,
as a function of the number of lattice spacings $d$ used to construct the
derivative with respect to $z_T $ in (\ref{ldiscrete}),(\ref{rdiscrete}).
Data are shown at a fixed $\hat{\zeta } =0.39$ and for three different
staple length parameters $\eta $; the isovector $u-d$ quark combination
was evaluated. The shown uncertainties are statistical jackknife errors.}
\label{slopeplot}
\end{figure}

A critical element in the evaluation of quark orbital angular
momentum is a stable extraction of the $z_T $-derivative
in (\ref{ldersingle}), as realized by the finite difference in
(\ref{rdiscrete}); a priori, one might expect strongly varying
behavior of the correlator $\Phi (z_T )$, reflecting the singular
nature of the $z_T \rightarrow 0$ limit. Such behavior could
potentially render the evaluation of quark orbital angular momentum
via (\ref{rdiscrete}) infeasible. Fig.~\ref{slopeplot} displays results
for the ratio (\ref{rdiscrete}) as a function of the distance $d$
used to construct the finite difference approximation to the
$z_T $-derivative, for a fixed Collins-Soper parameter $\hat{\zeta } $
and several fixed staple lengths $\eta |v|$; the derivative with
respect to $\Delta_{T} $ was evaluated as a finite difference as
discussed in the previous section. The data shown correspond to the
isovector $u-d$ quark channel, in which disconnected contributions
cancel and the renormalized number of valence quarks is $n=1$, i.e.,
the ratio (\ref{rdiscrete}) is directly a measure of quark orbital
angular momentum.

The ratio is fairly stable as $d$ is varied, implying an approximately
linear behavior of (the $\Delta_{T} $-odd part of) $\Phi (z_T )$ at small
$z_T $. In particular, the variation of the ratio data with $d$ tends to
decrease as one goes to smaller $d$ and one can assign an approximate value
to the $z_T $-derivative in (\ref{ldersingle}), as defined by the numerator
of (\ref{rdiscrete}), employing the data at small values of $d$. It should
be noted that, at the finite lattice spacing $a$ employed in the present
work, there is no clear separation between, on the one hand, the regime of
finite $d$, such that $|z_T |=da\rightarrow 0$ as $a\rightarrow 0$, in
which the numerator of (\ref{rdiscrete}) defines the $z_T $-derivative in
(\ref{ldersingle}); and, on the other hand, the regime of finite
physical $|z_T |=da$, in which (\ref{rdiscrete}) can be viewed as
defining a generalized quark orbital angular momentum, cf.~(\ref{rdisczt}).
The behavior of the data represents a mixture of the characteristics
of the two regimes. As $a\rightarrow 0$, a separation of the regimes
will be achieved; the behavior of the data in Fig.~\ref{slopeplot}
suggests that it may be plausible to expect that the ratio
(\ref{rdiscrete}) will indeed become sharply defined in the finite
$d$ regime, i.e., the different discretizations of the $z_T $-derivative
given by the different values of $d$ converge. On the other hand, in
view of Fig.~\ref{slopeplot} it appears reasonable to suppose that,
in the finite $|z_T |=da$ regime, the generalized quark orbital angular
momentum $(L_3 /n) (|z_T |)$ will exhibit a smooth limit towards the actual
quark orbital angular momentum as $z_T \rightarrow 0$. It will be useful
to explore and test these expectations in future work employing a variety
of lattice spacings.

In the discussion to follow, the results obtained at $d=1$ will be used
as the best available realization of the $z_T $-derivative in
(\ref{ldersingle}). Having thus settled on a definite scheme of
evaluating both of the derivatives in (\ref{ldersingle}), the remaining
parameters on which the results for the quark orbital momentum depend
are the staple length $\eta $ and the Collins-Soper parameter
$\hat{\zeta } $, where an extrapolation to large $\hat{\zeta } $
must be performed. Note that Fig.~\ref{slopeplot} provides a first
glimpse of the fact that data at different $\eta $ can be clearly
distinguished, and that their magnitude is enhanced at nonvanishing
$\eta $. This will be studied in detail further below. Before exploring
the dependence on the staple length $\eta $, it is useful to examine
the $\eta =0$ case, corresponding to Ji quark orbital angular momentum.

\begin{figure}
\centerline{\psfig{file=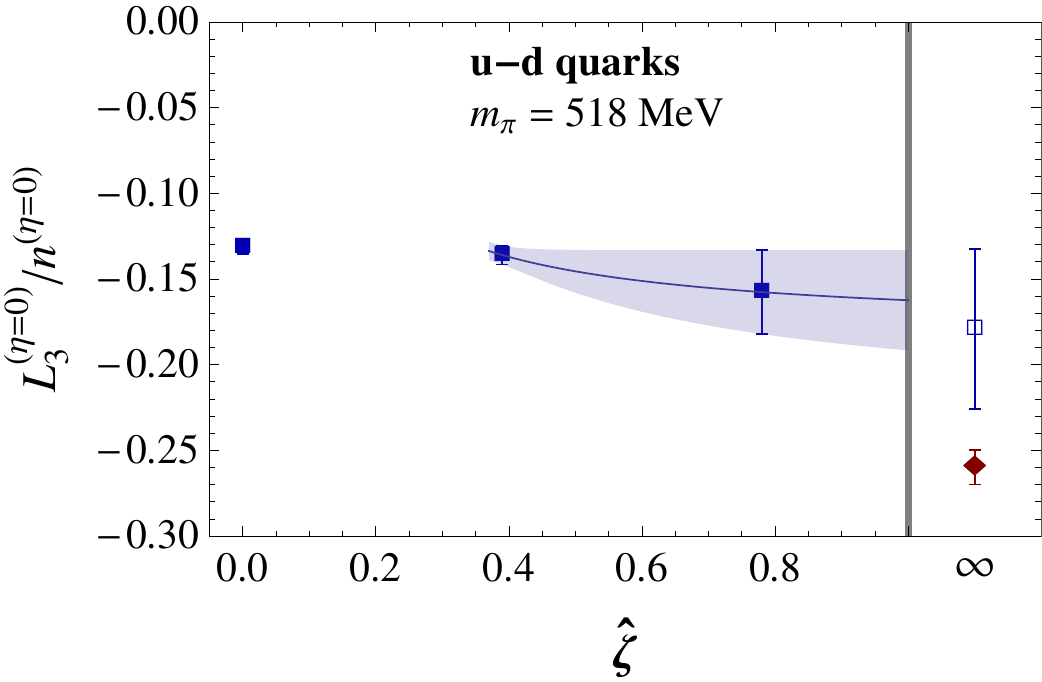,width=8.5cm} }
\caption{
Ji quark orbital angular momentum, i.e., the $\eta =0$ limit, as a
function of $\hat{\zeta } $, with an ad hoc extrapolation to infinite
$\hat{\zeta } $ (open square). The filled diamond represents the value
extracted on the same ensemble via Ji's sum rule. The isovector $u-d$
quark combination was evaluated. The shown uncertainties are statistical
jackknife errors.}
\label{jiplot}
\end{figure}

Fig.~\ref{jiplot} displays the isovector Ji quark orbital angular momentum
for all available values of $\hat{\zeta } $, together with an
extrapolation to infinite $\hat{\zeta } $ using the fit ansatz
$A+B/\hat{\zeta } $. It should be emphasized that this ad hoc ansatz,
which will be used repeatedly in the following, is not motivated by
any theoretical argument at this point, but has merely been observed
to provide a good fit of data in a previous investigation of the
Boer-Mulders TMD ratio \cite{bmlat}, for which an analogous extrapolation
in $\hat{\zeta } $ is performed. The extrapolation is thus intended to
give simply a heuristic idea of where the $\hat{\zeta } $ limit may
plausibly be located. The extrapolated value is confronted with the
one obtained on this same lattice ensemble from the standard evaluation
of quark orbital angular momentum via Ji's sum rule \cite{LHPC_2}.
While the sum rule value lies within two statistical standard deviations
of the extrapolated value, a genuine discrepancy between the two is in fact
expected, as discussed at the end of the previous section. The replacement
of the $\Delta_{T} $-derivative in (\ref{rdiscrete}) by a finite
difference using a rather large value of $|\Delta_{T} |$ represents
a substantial underestimate, possibly by up to a factor of two. The
discrepancy seen in Fig.~\ref{jiplot} is compatible with this
expectation. Indeed, if the extrapolated value in Fig.~\ref{jiplot}
were to be enhanced by a factor of two, it would overshoot the sum
rule value, but still be consistent with it within statistical
uncertainty. It thus does not seem implausible to expect that an
accurate estimate of the $\Delta_{T} $-derivative in future
numerical work will lead to consistent results for Ji quark orbital
angular momentum from the present direct evaluation method and the
standard sum rule evaluation. In order to roughly cancel the systematic
bias stemming from the underestimate of the $\Delta_{T} $-derivative
in the discussion of the $\eta \neq 0$ results below, data will be
presented relative to the $\eta =0$ Ji value.

\begin{figure}
\centerline{\psfig{file=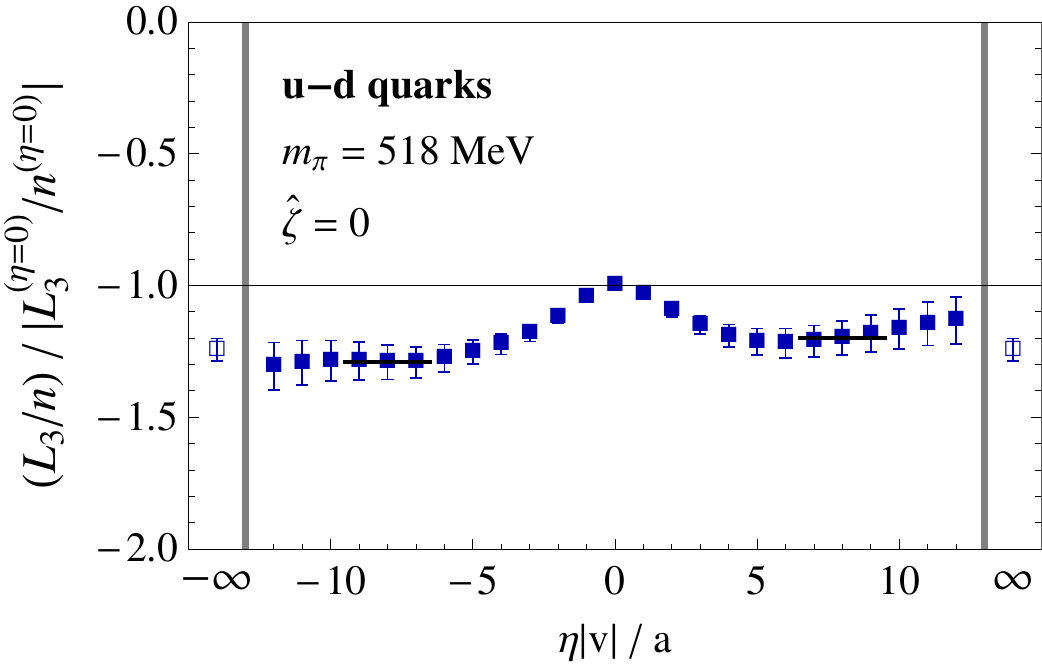,width=8.5cm} }
\centerline{\psfig{file=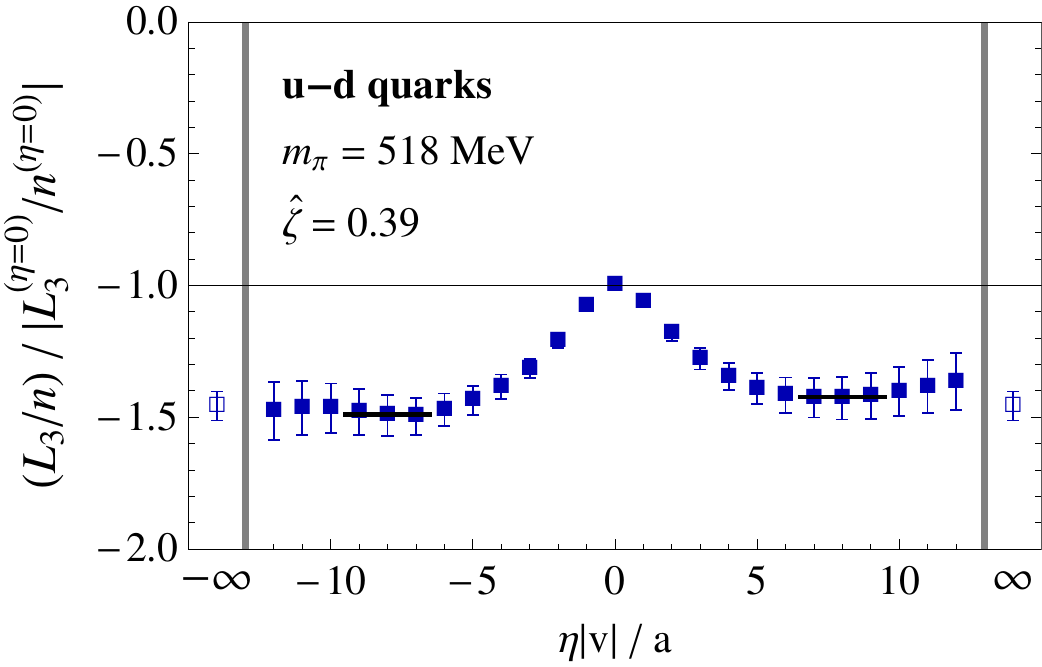,width=8.5cm} }
\centerline{\psfig{file=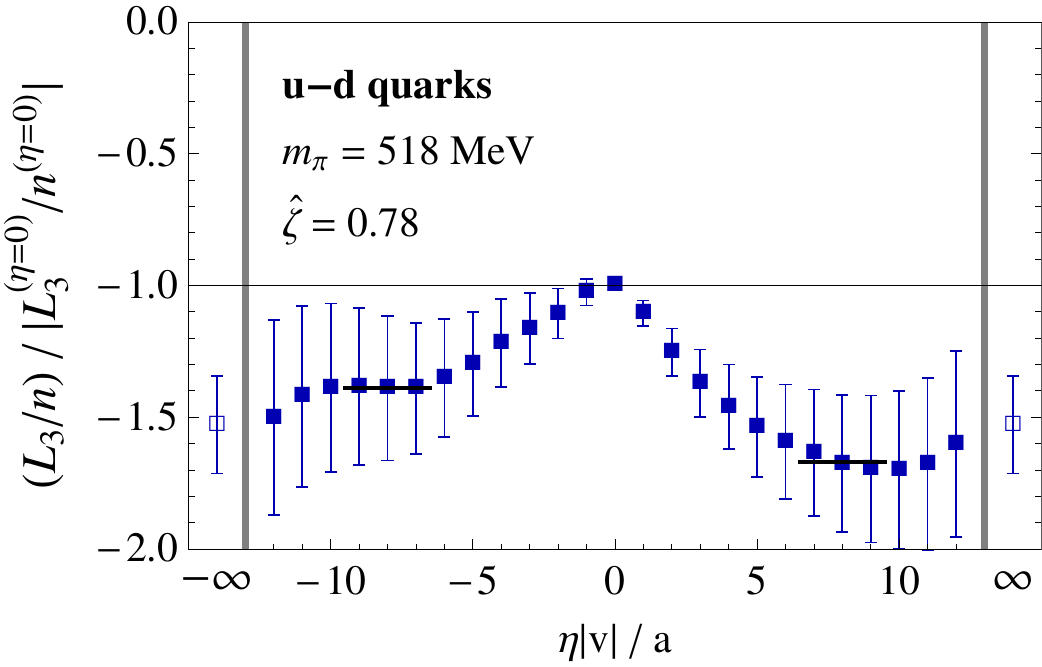,width=8.5cm} }
\caption{
Quark orbital angular momentum as a function of staple length parameter
$\eta $, normalized to the magnitude of the $\eta =0$ Ji value. The quark
orbital angular momentum is a time-reversal even quantity, i.e., even
under $\eta \rightarrow -\eta $. The extrapolated values are obtained by
averaging over the $\eta >0$ and $\eta <0$ plateau values, which are
each extracted by a fit to the $|\eta | |v|/a =7,\ldots ,9$ range.
The isovector $u-d$ quark combination was evaluated; the three panels
display the data for the three available values of $\hat{\zeta } $.
The shown uncertainties are statistical jackknife errors.}
\label{etaplot}
\end{figure}

The dependence of quark orbital angular momentum on the staple length
$\eta $ represents the most interesting feature of the present
investigation. As discussed in section~\ref{oamsec}, while the
$\eta =0$ case corresponds to Ji quark orbital angular momentum, the
$\eta \rightarrow \infty $ limit yields Jaffe-Manohar quark orbital
angular momentum, which, contrary to the former, has hitherto not been
accessible within Lattice QCD. The difference between the two can be
interpreted \cite{burk} as the integrated torque, due to the final state
interactions encoded in the staple-shaped gauge link, accumulated by the
struck quark in a deep inelastic scattering process as it is
leaving the proton. Since, in the present calculation, the staple
length can be varied in small steps, one in fact obtains a
quasi-continuous, gauge-invariant interpolation between the Ji and
Jaffe-Manohar limits. This is exhibited in Fig.~\ref{etaplot}, which
shows the variation of isovector orbital angular momentum with the staple
length parameter $\eta $; the three panels correspond to the three
available values of the Collins-Soper parameter $\hat{\zeta } $. In each
of the panels, following the change of orbital angular momentum as $\eta $
grows corresponds to observing the struck quark gather up torque on
its trajectory leaving the proton, until it reaches the Jaffe-Manohar
limit. Note that the data are displayed in units of (the magnitude of)
Ji quark orbital angular momentum, obtained at $\eta =0$. 

\begin{figure}
\centerline{\psfig{file=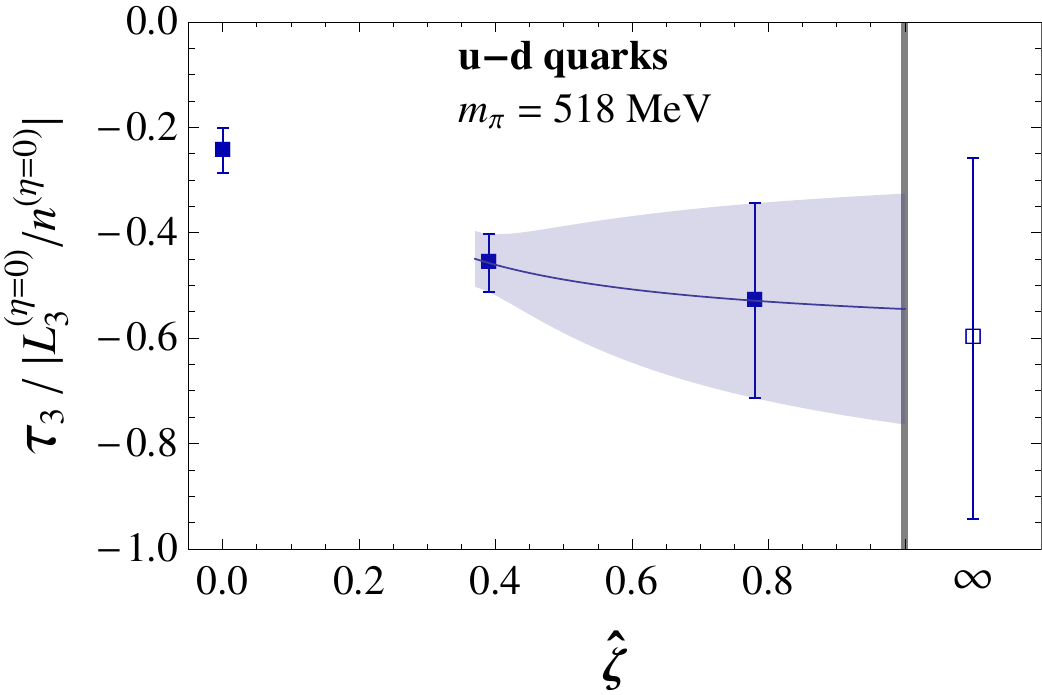,width=8.5cm} }
\caption{
Torque accumulated by the struck quark in a deep inelastic scattering
process, cf.~main text, normalized to the magnitude of the $\eta =0$
Ji orbital angular momentum, as a function of $\hat{\zeta } $, with an
ad hoc extrapolation to infinite $\hat{\zeta } $. The isovector $u-d$
quark combination was evaluated.  The shown uncertainties are
statistical jackknife errors.}
\label{torqueplot}
\end{figure}

The effect of the final state interactions seen in the data is substantial.
It can be clearly resolved even within the limited statistics employed
in the present exploratory calculation, and it enhances quark orbital
angular momentum compared with the initial Ji value. As seen in
Fig.~\ref{etaplot}, the effect increases with rising Collins-Soper
parameter $\hat{\zeta } $, approaching magnitudes such that the
integrated torque accumulated in reaching the Jaffe-Manohar limit amounts
to as much as roughly half of the initial Ji orbital angular momentum
carried by the quark. The increasing behavior with $\hat{\zeta } $
also implies that the effect is likely to survive extrapolation to
the $\hat{\zeta } \rightarrow \infty $ limit. A corresponding
extrapolation, again employing the fit ansatz $A+B/\hat{\zeta } $,
is exhibited in Fig.~\ref{torqueplot}. Shown are data for the integrated
torque
\begin{equation}
\tau_{3} = \frac{L^{(\eta = \infty )}_{3} }{n^{(\eta = \infty )} }
-\frac{L^{(\eta = 0)}_{3} }{n^{(\eta = 0)} } \ ,
\label{torquedef}
\end{equation}
i.e., the difference between Jaffe-Manohar and Ji quark orbital momentum,
in units of the magnitude of Ji quark orbital momentum. A signal is
indeed obtained in the $\hat{\zeta } \rightarrow \infty $ limit. Note
that the integrated torque $\tau_{3} $ corresponds to a Qiu-Sterman type
correlator \cite{burk,eomlir}. The calculation of $\tau_{3} $ presented
here amounts to an evaluation of that genuine twist-three term.

\begin{figure}
\centerline{\psfig{file=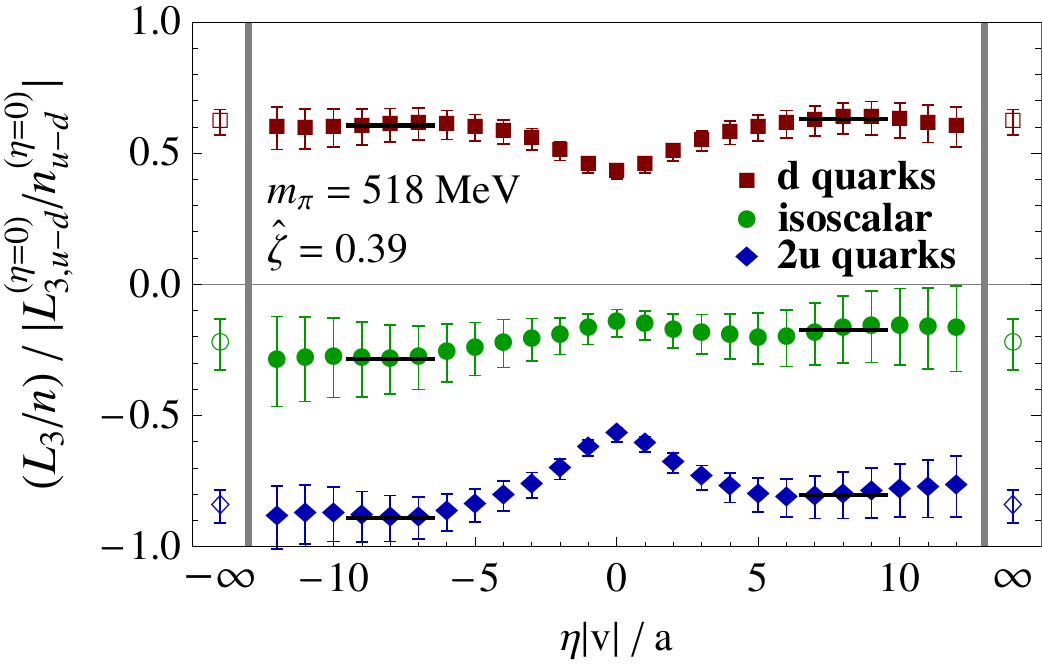,width=8.5cm} }
\caption{
Flavor-separated orbital angular momentum as a function of staple
length parameter $\eta $, analogous to Fig.~\ref{etaplot}, at fixed
$\hat{\zeta } =0.39$. Shown are data for $d$ quarks, for two $u$ quarks
(i.e., the $u$-quark data for $L_3^U /n$ have been multiplied by $2$ to
compensate for $n=2$ in the $u$-quark case), and the isoscalar total
quark orbital angular momentum. The latter was obtained by adding the
``$d$'' and ``$2u$'' data. All data are still normalized by the magnitude
of $u-d$ Ji orbital angular momentum, i.e., at $\eta =0$, the ``$2u$''
and ``$d$'' data differ by unity. The shown uncertainties are statistical
jackknife errors.}
\label{flav_etaplot}
\end{figure}

\begin{figure}
\centerline{\psfig{file=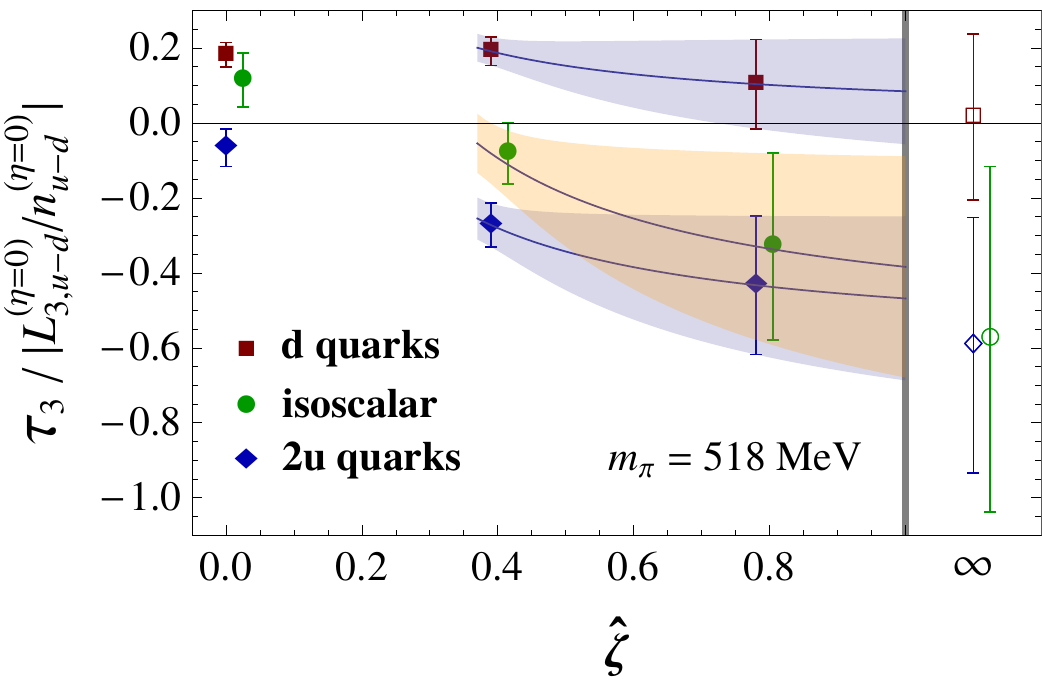,width=8.5cm} }
\caption{Flavor-separated torque accumulated by the struck quark in a
deep inelastic scattering process, as a function of $\hat{\zeta } $,
analogous to Fig.~\ref{torqueplot}, with ad hoc extrapolations to
infinite $\hat{\zeta } $. Shown are data for $d$ quarks, for two
$u$ quarks (i.e., the $u$-quark data for $\tau_{3} $ have been
multiplied by $2$ to compensate for $n=2$ in the $u$-quark case),
and the isoscalar combination. The latter
was obtained by adding the ``$d$'' and ``$2u$'' data. All data are
still normalized by the magnitude of $u-d$ Ji orbital angular momentum.
Isoscalar data are slightly displaced horizontally for better visibility.
The shown uncertainties are statistical jackknife errors.}
\label{flav_torqueplot}
\end{figure}

All data presented above concern the isovector $u-d$ quark combination,
in which disconnected diagrams cancel. Of interest is also the decomposition
by flavor, shown in Figs.~\ref{flav_etaplot} and \ref{flav_torqueplot}.
In these flavor-decomposed data, disconnected contributions are omitted;
they are, however, expected to be minor at the heavy pion mass used for
the present calculation. Fig.~\ref{flav_etaplot} exhibits quark orbital
angular momentum as a function of staple length parameter $\eta $, for a
fixed value of the Collins-Soper parameter $\hat{\zeta } $, analogous to
Fig.~\ref{etaplot}. Displayed are data for the $d$ quark and for the
two $u$ quarks, i.e., $L_3^U /n$ for $u$ quarks has been multiplied by 2
to compensate for $n=2$ in the $u$ quark case. Displayed furthermore is
the total (isoscalar) quark orbital angular momentum, which was obtained
here simply by adding the aforementioned ``$d$'' and ``$2u$'' contributions
(at finite statistics, this may differ slightly from evaluating instead
$3L_{3,u+d} /n_{u+d} $). The data are again given in units of the
magnitude of the isovector Ji quark orbital angular momentum, i.e.,
in Fig.~\ref{flav_etaplot} at $\eta =0$, the difference between the
``$2u$'' and ``$d$'' data is unity.

The flavor-separated data in Fig.~\ref{flav_etaplot} reproduce the
well-known cancellation between the $d$- and $u$-quark orbital angular
momenta in the proton \cite{LHPC_1,LHPC_2}, which combine to yield only
a small negative residual contribution to the spin of the proton at the
pion mass used in the present study. This property persists as one
departs from Ji quark orbital momentum and adds torque to arrive
at Jaffe-Manohar quark orbital momentum. Fig.~\ref{flav_torqueplot}
displays results for the integrated torque, cf.~(\ref{torquedef}), in
analogy to Fig.~\ref{torqueplot}, including extrapolations of the
data at different Collins-Soper parameters $\hat{\zeta } $ to the
$\hat{\zeta } \rightarrow \infty $ limit. The results at
$\hat{\zeta } =0.78$ exhibit fairly large statistical fluctuations,
which may be the source of the seemingly uncharacteristic behavior
of the $d$-quark data point. It would be natural to expect monotonous
behavior as a function of $\hat{\zeta } $, and the aforementioned
data point may thus well represent a downward fluctuation; it leads
to an extrapolated $d$-quark torque compatible with zero, driving
the isoscalar combination to negative values (albeit with a large
statistical uncertainty). Another, more physical, reason to suspect a
spurious downward fluctuation in this case is that final state interaction
effects on quark transverse momenta in the proton generally tend to be
stronger for a $d$-quark than a (single) $u$-quark, as embodied, e.g.,
in the Sivers shift or the Boer-Mulders shift \cite{tmdlat,bmlat}
This suggests that a subtle interplay between transverse positions and
momenta of the quarks would be needed to generate a deviating pattern
for the integrated torque. Indeed, the substantial statistical
uncertainties at $\hat{\zeta } =0.78$ do in fact still render the data
compatible with a $d$-quark integrated torque that is stronger than a
(single) $u$-quark integrated torque. A higher statistics calculation is
needed to draw definite conclusions on this point. At present, in view of
the sizeable uncertainties, the extrapolated results are nevertheless
compatible with the conclusions drawn above from Fig.~\ref{flav_etaplot}
at the fixed value $\hat{\zeta } =0.39$.

\section{Conclusions and outlook}
The present exploration demonstrates the feasibility of calculating
the quark orbital angular momentum in a proton in Lattice QCD directly,
using a Wigner distribution-based approach. Given a Wigner distribution of
quark parton positions and momenta in the transverse plane with respect
to proton motion, one can directly evaluate their average cross product,
i.e., longitudinal orbital angular momentum. The Wigner
distribution in question can be derived from appropriate proton matrix
elements \cite{lorce}. As is familiar from the definition of transverse
momentum-dependent parton distributions (TMDs), transverse momentum
information can be extracted by employing a quark bilocal operator;
the transverse quark separation in this operator is Fourier conjugate to the
quark transverse momentum. On the other hand, by generalizing the TMD
matrix element to non-zero momentum transfer, the TMD information
is supplemented with impact parameter information. One thus arrives at
generalized transverse momentum-dependent parton distributions (GTMDs)
\cite{mms}, which, appropriately Fourier-transformed, yield the desired
Wigner distribution. In view of this construction, many of the concepts and
techniques employed in lattice TMD studies \cite{tmdlat,bmlat} can be
brought to bear on the direct evaluation of quark orbital angular momentum
in a proton.

In order to define a gauge-invariant observable, the bilocal TMD operator
must contain a gauge connection between the quarks. Different choices of
the path along which the connection is evaluated encode different
definitions of quark orbital angular momentum \cite{jist,hatta,burk},
including the Ji and Jaffe-Manohar definitions. The approach developed here
has thus opened the possibility of going beyond the lattice studies of quark
orbital angular momentum carried out to date, which were limited to
calculating specifically Ji orbital angular momentum via Ji's sum rule.
Indeed, the present investigation has provided a quasi-continuous,
gauge-invariant interpolation between Ji and Jaffe-Manohar orbital
angular momentum, by deforming the gauge connection path in small steps.
The difference between the two definitions is given by Burkardt's torque
\cite{burk}, i.e., the integrated torque accumulated by the struck quark
in a deep inelastic scattering process as it is leaving the proton; it
corresponds to a genuine twist-three, Qiu-Sterman type correlator. The
aforementioned interpolation allows one to directly follow the gradual
accumulation of this torque along the quark's path.

The gauge connection contained in the bilocal TMD operator is associated
with ultraviolet divergences which, in the continuum Collins factorization
scheme \cite{aybat,collbook,spl2}, are absorbed into multiplicative soft
factors appearing alongside the renormalization factors associated with the
quark fields. As in lattice TMD studies, also in the present context an
appropriate ratio can be formed in which these factors cancel; it is
assumed that their multiplicative nature carries over into the lattice
formulation. The renormalized quantity evaluated in effect is $L_3 /n$,
the longitudinal quark orbital angular momentum component in units of the
number of valence quarks, for the quark flavor under consideration. Accessing
this quantity entails, in particular, evaluating the derivative of the
relevant proton matrix element with respect to the quark separation $z_T $
in the employed quark bilocal operator, in the limit of vanishing $z_T $.
This limit must be taken with care owing to the associated ultraviolet
singularities. The behavior of the gathered data with varying $z_T $ was
considered in particular, revealing a fairly stable behavior that suggests
that the aforementioned $z_T $-derivative can be extracted unambiguously
in the continuum limit. This is a prerequisite for the application of
the method.

The principal new physical insight gleaned from the present exploration
concerns the behavior of Jaffe-Manohar quark orbital angular momentum
relative to its Ji counterpart. Their difference, Burkardt's torque,
is clearly resolvable in the data and is sizeable, amounting to roughly
one half of Ji quark orbital angular momentum. The torque enhances
Jaffe-Manohar quark orbital angular momentum relative to Ji quark orbital
angular momentum. As already noted above, the data gathered within this
investigation provide a quasi-continuous, gauge-invariant interpolation
between the two limits.

The most significant shortcoming of the data set obtained in the present
exploration is the limited set of proton momenta, both in the longitudinal
and the transverse directions. In particular, only one fairly large
value of the momentum transfer $\Delta_{T} $ in the transverse plane is
available, and the $\Delta_{T} $-derivative necessary to extract orbital
angular momentum is substantially underestimated by the finite difference
obtained using the aforementioned $\Delta_{T} $ value. Thus, while the
above conclusions concerning the relative comparison between Jaffe-Manohar
and Ji quark orbital angular momentum are presumably robust with respect
to this systematic bias, in absolute terms, the results extracted
systematically underestimate the quark orbital angular momentum present
in the proton. This, indeed, is manifested clearly in the case of Ji
orbital angular momentum, through the comparison with the corresponding
indirect Ji sum rule lattice evaluation on the same ensemble. Removing
this systematic bias is foremost on the agenda for future work. A
calculation is in preparation which will incorporate a direct method to
evaluate the $\Delta_{T} $-derivative exactly \cite{rome,nhasan,smpriv},
related to twisted boundary conditions on the quark fields. The
aforementioned comparison with the standard evaluation of Ji orbital
angular momentum via Ji's sum rule will provide a valuable benchmark
for the removal of systematic bias.

Moreover, the data are to be extrapolated to large Collins-Soper
parameter $\hat{\zeta } $, realized in the employed Lorentz frame
via large proton momenta. Although the availability of several
longitudinal proton momenta in the gathered data set allowed for an
ad hoc estimation of the large-$\hat{\zeta } $ limit, calculations at
larger proton momenta are clearly desirable. In this respect,
the improved proton sources explored in \cite{highmom,sspriv} provide
a perspective for generating data that permit a more quantitative
treatment of the large-$\hat{\zeta } $ limit.

Further topics for future exploration include the investigation of varying
lattice spacing $a$, with a view towards studying, in particular, the
$a\rightarrow 0$ behavior of the $z_T $-derivative discussed above, as
well as observing the evolution of the extracted quark orbital angular
momentum. Also, efforts to approach the physical pion mass in these
calculations must be undertaken. On a more theoretical level, a decomposition
of the proton matrix element under consideration into Lorentz-invariant
amplitudes, analogous to the one employed in lattice TMD studies
\cite{tmdlat} ought to provide valuable further insight into the
systematics of the data and aid in the extrapolation to large
$\hat{\zeta } $, cf.~the study of the Boer-Mulders TMD ratio \cite{bmlat}.

\section*{Acknowledgments}
This work benefited from fruitful discussions with M.~Burkardt, W.~Detmold,
J.~Green, R.~Gupta, S.~Liuti, C.~Lorc\'{e}, S.~Meinel, B.~Musch, J.~Negele,
S.~Syritsyn and B.~Yoon. The lattice calculations performed in this work
relied on code developed by B.~Musch, as well as the Chroma software
suite \cite{chroma}, and employed computing resources provided by
the U.S.~DOE through USQCD at Jefferson Lab. Support by the U.S.~DOE
through grant DE-FG02-96ER40965 as well as through the TMD Topical
Collaboration is acknowledged.

\end{document}